\documentstyle[epsfig,12pt]{article}

\newcommand{\Z}{{\sf Z \!\!\! Z}}

\newcommand{\Sign}{\mbox{Sign}}
\newcommand{\PP}{\overline{\Psi} \Psi}
\makeatletter
\@addtoreset{equation}{section}
\makeatother

\title{Meron-cluster algorithms and chiral symmetry breaking
in a $(2+1)$-d staggered fermion model
\footnote{This work is supported in part by funds provided by the U.S.
Department of Energy (D.O.E.) under cooperative research agreement
DE-FC02-94ER40818 and by the Schweizerischer Nationalfonds.}}
\author{J. Cox$^{{\rm a}}$ and K. Holland$^{{\rm b}}$ \\ \\
$^{{\rm a}}$ Center for Theoretical Physics, \\
Laboratory for Nuclear Science, and Department of Physics \\
Massachusetts Institute of Technology (MIT) \\
Cambridge, Massachusetts 02139, U.S.A. \\ \\
$^{{\rm b}}$ Institute for Theoretical Physics \\
University of Bern \\
Sidlerstrasse 5, CH-3012 Bern, Switzerland \\ \\
MIT-CTP 2965, BUTP-2000-08 \\ \\}
 
\begin{document}

\maketitle
\begin{abstract} \normalsize
 
The recently developed Meron-Cluster algorithm completely solves the
exponentially difficult sign problem for a number of models previously
inaccessible to numerical simulation. We use this algorithm in a
high-precision study of a model of $N=1$ flavor of staggered fermions in
(2+1)-dimensions with a four-fermion interaction. This model cannot be
explored using standard algorithms. We find that the $\Z(2)$ chiral symmetry
of this model is spontaneously broken at low temperatures and that the     
finite-temperature chiral phase transition is in the universality class of the
2-d Ising model, as expected.
\end{abstract}
 
\maketitle
 
\newpage

\section{Introduction}

There are a number of models of interest which suffer from a very severe sign
problem. This includes QCD and other field theories with a non-zero chemical
potential or a non-zero vacuum angle or odd numbers of fermion flavors, frustrated
quantum spin systems, like the quantum antiferromagnet in an external magnetic
field, and models for strongly-correlated electrons, like the Hubbard model for 
high-temperature superconductivity. These models have a Boltzmann weight
which can be negative or even complex and so cannot be interpreted as a 
probability. This difficulty can be overcome in numerical simulations by
including the sign or phase of the Boltzmann weight with observables. 
Unfortunately, this leads to large cancellations and gives exponentially small
observables. This requires exponentially large statistics, which makes it in 
practice impossible to simulate these models numerically.

Recently, a new technique has been developed, called Meron-Cluster algorithms
\cite{Cha99a}, which completely solves the sign problem for some of these
models \cite{Cha99b,Cox99}. It identifies the origin of the sign problem with
properties of the clusters, which enables it to be eliminated. Cluster
algorithms in general are extremely efficient at exploring configuration
spaces and very often do not suffer from critical slowing down as a phase 
transition is approached, unlike many other algorithms. For example, in this 
and previous papers, we can work directly in the chiral limit with massless
fermions. Combined with the ability to construct improved estimators, we can
perform a high precision study of these models with only modest statistics. 

In this paper, we explore a model of $N=1$ flavor of staggered fermions in
(2+1)-dimensions with a four-fermion interaction. This model has a very severe
sign problem and cannot be simulated with standard techniques. We build a
Meron-Cluster algorithm, which we use to perform a high-precision study. We
find that the $\Z(2)$ chiral symmetry of this model is spontaneously broken at
low temperatures and, using finite-size scaling analysis, we verify that the
finite-temperature chiral phase transition is in the universality class of the
2-d Ising model. A recent study of the same model with
$\Z(2)$ chiral symmetry in (3+1)-dimensions has shown, using a Meron-Cluster
algorithm, that a finite temperature chiral phase transition occurs which has
the universal behavior of the 3-d Ising model \cite{Cha99b}. The work
presented in this paper concerns a different universality class and also
constructs more observables than were previously considered.

The identification of the finite temperature critical behavior is not entirely
straightforward \cite{Kog98}. A model of $N$ fermion flavors with a
four-fermion interaction shows mean-field behavior in the $N = \infty$ 
limit. On the other hand, at finite $N$ one finds the non-trivial critical 
behavior that one expects based on dimensional reduction and standard 
universality arguments. For example, in
\cite{Kog99} it has been verified that the chiral phase transition in a
$(2+1)$-d four-fermion interaction model with $N = 4$ flavors and $\Z(2)$
chiral symmetry is in the universality class of the 2-d Ising model. Due to
the fermion sign problem, standard fermion simulation methods often do not
work in models with too small a number of flavors. The work presented in this
paper shows that the same universal behavior holds for $N=1$ flavor. 

The standard technique to deal with fermions in Monte-Carlo simulations is to
integrate them out, resulting in a non-local bosonic theory. The Meron-Cluster
algorithm does not integrate out the fermions, instead we
describe them with a local theory using a Fock space basis of occupation
number. The fermion sign arises as a non-local property due to the permutation
of fermion world lines. Using probabilistic rules, we connect neighboring
lattice sites, producing closed loops, which are the clusters. A cluster is
flipped by making all of its occupied sites empty and its empty ones
occupied. Such a cluster flip can change the fermion sign by changing the
permutation of fermion world lines. A cluster whose flip changes the sign we
call a meron. We can tell if a cluster is a meron simply from its structure. A
typical configuration contains many merons, yet the observables of interest
are only non-zero for configurations with very few or no merons. The signals
from standard Monte Carlo algorithms are so exponentially small because
the Markov chain explores a vast configuration space, yet only an
exponentially small sub-space makes {\em any} contribution to measurables. By
restricting ourselves to only explore the relevant sub-space, we completely
solve the sign problem.

This paper is organized as follows. In section 2, we present the fermionic
model which we have studied, calculate its partition function using the
Hamiltonian formulation and find that there is a sign problem. In section 3,
we describe briefly the Meron-Cluster algorithm which we have used to perform 
numerical simulations of this model. We present the results of the simulations
in section 4 and give our conclusions in section 5. 

\section{The Staggered Fermion Model}

We consider staggered fermions in the Hamiltonian formulation on a
2-dimensional spatial lattice of extent $L$, which is even. 
The Hamiltonian operator is 
\begin{eqnarray}
H &=& \sum_{x,i} h_{x,i} + m \sum_x (-1)^{x_1 + x_2} \Psi^+_x \Psi_x
\nonumber \\
h_{x,i} &=& \eta_{x,i} (\Psi^+_x \Psi_{x+\hat{i}} + \Psi^+_{x+\hat{i}} \Psi_x)
+ G (\Psi^+_x \Psi_x - \frac{1}{2})(\Psi^+_{x+\hat{i}} \Psi_{x+\hat{i}} -
\frac{1}{2}),
\end{eqnarray}
where $\eta_{x,1}=1$ and $\eta_{x,2}=(-1)^{x_1}$ are the standard
Kawamoto-Smit phases for staggered fermions and $G$ is a 
constant. The fermionic operators satisfy the usual anticommutation relations
$\{\Psi_x,\Psi_y\} = \{\Psi^+_x,\Psi^+_y\} = 0, \; \{\Psi^+_x,\Psi_y\} =
\delta_{xy}$. The same model in (3+1)-dimensions was explored in
\cite{Cha99b}. We refer the reader to this paper, where various features of
the model and the Meron-Cluster algorithm are discussed in more detail than we
give here.

In the Hamiltonian formulation of the theory, fermion doubling on the lattice
occurs only in the spatial dimensions. Using staggered fermions, the Dirac
components of a spinor are distributed spatially, reducing the number of
fermion flavors by a factor of four. Thus this (2+1)-dimensional model contains
$N=1$ fermion flavor. The model has a global $U(1)$ symmetry corresponding
to conserved particle number, as the total particle number operator commutes
with the Hamiltonian
\begin{equation}
N = \sum_x \Psi^+_x \Psi_x, \; [H,N] = 0.
\end{equation}
Furthermore the Hamiltonian has, for $m=0$, a discrete $\Z(2)$ symmetry
corresponding to shifts by one lattice spacing. However the mass term breaks
that symmetry explicitly. For a single flavor of massless fermions, the
symmetry of the lattice model is $U(1) \otimes \Z(2)$ and we refer to
the discrete symmetry as chiral symmetry. In the continuum, a single massless
fermion flavor has a $U(1)$ axial symmetry (there is no gauge interaction, so
this symmetry is not anomalously broken). The discrete $\Z(2)$ symmetry is the
lattice remnant of this continuous symmetry. From now on, we set $m=0$ and
explore the behavior of the chiral symmetry of this model. The symmetries of
staggered fermions are discussed in detail in Ref. \cite{Sus77}. If the
$\Z(2)$ chiral symmetry is spontaneously broken at some finite temperature, 
from universality we expect this to be a second-order phase transition. As the
critical point is approached, the correlation length $\xi$ diverges and the
system becomes insensitive to the time extent. Due to dimensional reduction,
we expect a finite-temperature chiral phase transition in this model to belong
to the 2-d Ising universality class.   

The partition function of the model is
\begin{eqnarray}
&& Z = \mbox{Tr} [\exp(-\beta H)] = \lim_{M \rightarrow \infty} \mbox{Tr}
[\exp(-\epsilon H)]^M \\ \nonumber
&& = \lim_{M \rightarrow \infty} \mbox{Tr} [\exp(-\epsilon H_1)
\exp(-\epsilon H_2) \exp(-\epsilon H_3) \exp(-\epsilon H_4)]^M,
\end{eqnarray}
where we use the Suzuki-Trotter decomposition to divide the Euclidean time
extent $\beta$ into $4 M$ time slices, the lattice spacing in the time
direction being $\epsilon = \beta/M$. The Hamiltonian operator is decomposed
into four parts $H = H_1 + H_2 + H_3 + H_4$. All of the terms that contribute
to a particular $H_i$ commute with one another, as each term is an interaction
between nearest-neighbors and each lattice site appears in only one such 
nearest-neighbor pair. However, the $H_i$ do not commute with one another. We 
note that it is not actually necessary to discretize the time direction, as it
is possible to work directly in the Euclidean time continuum \cite{Bea96}. 

We can equivalently describe this model with bosonic operators, using a
transformation by Jordan and Wigner \cite{Jor28}. We order the lattice sites
on each time slice arbitrarily into a chain, which can be done in any number
of spatial dimensions. For example, a possible ordering of points in two
spatial dimensions is by an index $l = x_1 + (x_2-1) L$. The fermionic
operators are now represented by a chain of Pauli matrices
\begin{eqnarray}
&& \Psi^+_x = \sigma^3_1 \sigma^3_2 ... \sigma^3_{l-1} \sigma^+_l, \;
\Psi_x = \sigma^3_1  \sigma^3_2 ... \sigma^3_{l-1} \sigma^-_l, \;
\Psi^+_x \Psi_x = \frac{1}{2} (\sigma^3_l + 1) \\ \nonumber
&& \sigma^{\pm} = \frac{1}{2} (\sigma^1 \pm i \sigma^2), \;
[\sigma^i_l,\sigma^j_m] = 2 i \delta_{lm} \epsilon^{ijk} \sigma^k_l,
\end{eqnarray}
where the spatial position $x$ is denoted by the index $l$ and the Pauli matrices
satisfy the usual commutation relations. To calculate the partition function of
the theory, we use the Fock space basis of occupation number $n_x = 0,1$ i.e.
the eigenstates of $\sigma^3$. The occupied and empty states are respectively
$|1\rangle$ and $|0\rangle$, which satisfy $\sigma^3 |1\rangle = |1\rangle$
and $\sigma^3 |0\rangle = - |0\rangle$. 

The time evolution operator $\exp(-\epsilon H_i)$ acts on a time slice of
occupation number states, producing the next time slice. This is decomposed
into the product of operators $\exp(-\epsilon h_{x,i})$ acting on
nearest-neighbor occupation states. The transfer matrix is
\begin{equation}
\label{transfer}
\exp(- \epsilon h_{x,i}) = \exp(\frac{\epsilon G}{4})
\left(\begin{array}{cccc}
\exp(- \frac{\epsilon G}{2}) & 0 & 0 & 0 \\ 
0 & \cosh \frac{\epsilon}{2} & \Sigma \sinh \frac{\epsilon}{2} & 0 \\ 
0 & \Sigma \sinh \frac{\epsilon}{2} & \cosh \frac{\epsilon}{2} & 0 \\ 
0 & 0 & 0 & \exp(- \frac{\epsilon G}{2}) \end{array} \right),
\nonumber \\ \
\end{equation}
the basis being $|00\rangle,|01\rangle,|10\rangle$ and $|11\rangle$, where
e.g. $|01\rangle$ represents state $|0\rangle$ at $x$ and $|1\rangle$ at
$x+\hat{i}$. If these nearest-neighbors are labelled $l$ and $m$, the 
off-diagonal transfer matrix elements have a factor $\Sigma = \eta_{x,i} 
\sigma^3_{l+1} \sigma^3_{l+2} ... \sigma^3_{m-1}$. Note that this operator is 
diagonal in the occupation number basis.

The partition function of the theory is given as a path integral
\begin{equation}
Z_f = \sum_n \Sign[n] \exp(- S[n]),
\end{equation}
where we sum over all possible configurations of occupation numbers $n(x,t) =
0,1$ on a $(2+1)$-d space-time lattice of points $(x,t)$. The Boltzmann factor
$\exp(- S[n])$ for a configuration is the product of the Boltzmann factors for
each space-time plaquette $\exp(-s[n(x,t),n(x+\hat i,t),n(x,t+1),n(x+\hat
i,t+1)])$, which are
\begin{eqnarray}
\label{Boltzmann}
&&\exp(- s[0,0,0,0]) = \exp(- s[1,1,1,1]) = \exp(- \frac{\epsilon G}{2}),
\nonumber \\
&&\exp(- s[0,1,0,1]) = \exp(- s[1,0,1,0]) = \cosh \frac{\epsilon}{2},
\nonumber \\
&&\exp(- s[0,1,1,0]) = \exp(- s[1,0,0,1]) = \sinh \frac{\epsilon}{2}.
\end{eqnarray}
All other plaquettes are illegal and have Boltzmann weight zero, as they 
represent non-conservation of fermion number. Any configuration which contains
illegal plaquettes has itself Boltzmann weight zero and makes no contribution
to the partition function. We are only interested in legal configurations,
which have to satisfy several constraints. Note that here we have dropped the 
overall factor $\exp(\epsilon G/4)$ that appeared in eq.(\ref{transfer}). The 
sign of a configuration, $\Sign[n]$, is also a product of space-time plaquette
contributions 
$\mbox{sign}[n(x,t),n(x+\hat i,t),n(x,t+1),n(x+\hat i,t+1)]$ with
\begin{eqnarray}
&&\mbox{sign}[0,0,0,0] = \mbox{sign}[0,1,0,1] = \mbox{sign}[1,0,1,0] = 
\mbox{sign}[1,1,1,1] = 1, \nonumber \\
&&\mbox{sign}[0,1,1,0] = \mbox{sign}[1,0,0,1] = \Sigma.
\end{eqnarray}
The occupied lattice sites define world-lines of fermions, which close due to
the periodicity of the Euclidean time direction. The world-lines are free to 
permute during their time evolution as the fermions interchange position and
each configuration has a well-defined permutation of fermions. The Pauli
exclusion principle tells us that the sign of a configuration is the
permutation sign of the fermions, hence $\Sign[n]=\pm1$. This non-local effect
is contained in the factors $\Sigma$ of each space-time plaquette.

The expectation value of a fermionic observable $A[n]$ is given by
\begin{eqnarray}
&& \langle A \rangle_f = \frac{1}{Z_f} \sum_n A[n] \Sign[n] \exp(-S[n])
= \frac{\langle A \Sign \rangle}{\langle \Sign \rangle}, \\ \nonumber
&& \langle \Sign \rangle = \frac{1}{Z_b} \sum_n \Sign[n] \exp(-S[n]), 
\end{eqnarray}
where $\langle ... \rangle$ means a measurement made in the bosonic ensemble,
whose partition function is $Z_b = \sum_n \exp(-S[n])$. To measure one
fermionic observable requires two bosonic measurements. The quantities of
physical interest which we measure are the chiral condensate $\PP$, the chiral 
susceptibility $\chi$ and a Binder cumulant $U$ of the chiral condensate, 
respectively
\begin{eqnarray}
&& \hspace{-1.5in} \PP[n] = \frac{\epsilon}{4} \sum_{x,t} (-1)^{x_1 + x_2} 
(n(x,t) - \frac{1}{2}), \nonumber \\
\chi = \frac{1}{\beta V} \langle (\overline{\Psi} \Psi)^2 
\rangle_f, &&
U = 1 - \frac{\langle (\PP)^4 \rangle_f}{3 [\langle (\PP)^2 \rangle_f]^2}.
\end{eqnarray}

\section{The Meron-Cluster Algorithm}

We now describe briefly the Meron-Cluster algorithm which we used to sample
the bosonic ensemble corresponding to the fermionic model without the sign
factor. We set $G=1$, for which the bosonic model is the isotropic
antiferromagnetic quantum Heisenberg model, whose Hamiltonian is $H = 
\sum_{x,i} (S_x^1 S_{x+\hat i}^1 + S_x^2 S_{x+\hat i}^2 + S_x^3 
S_{x+\hat i}^3)$, where $S_x^i = \frac{1}{2} \sigma_l^i$ is a spin $1/2$ operator
at the lattice site $x$, labelled by $l$ in the Jordan-Wigner chain. There
already exist extremely efficient cluster algorithms to simulate bosonic
quantum spin systems \cite{Wie92,Eve93,Eve97}, and the first cluster algorithm
for lattice fermions was constructed in \cite{Wie93}. These algorithms can be
implemented directly in the time continuum \cite{Bea96}, i.e. the
Suzuki-Trotter time discretization is not even necessary. In this study, we 
discretize the time direction.

\begin{table}[t]
% space before first and after last column: 1.5pc
% space between columns: 3.0pc (twice the above)
%\setlength{\tabcolsep}{1.5pc}
% -----------------------------------------------------
% adapted from TeX book, p. 241
\newlength{\digitwidth} \settowidth{\digitwidth}{\rm 0}
\catcode`?=\active \def?{\kern\digitwidth}
% -----------------------------------------------------
  \begin{center}
    \leavevmode
%    \tiny
    \begin{tabular}{|c|c|c|}
        \hline
        weight&configuration&break-ups\\
        \hline\hline
        $\exp\left(-\frac{\epsilon}{2}\right)$
        &\hspace{0.4cm}\begin{minipage}[c]{2.5cm}
                \epsfig{file=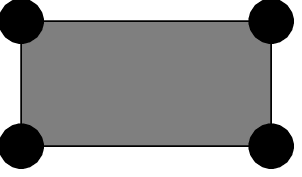,width=1.5cm,angle=270,
                bbllx=-5,bblly=0,bburx=55,bbury=85}
        \end{minipage}
        &\hspace{0.4cm}\begin{minipage}[c]{2.5cm}
                \vbox{\begin{center}
                \epsfig{file=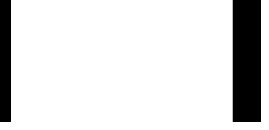,width=1.5cm,angle=270,
                bbllx=-20,bblly=0,bburx=55,bbury=85} 

                \hspace{-0.2cm}1
                \end{center}}
        \end{minipage}
        \\
        \hline
        $\cosh\left(\frac{\epsilon}{2}\right)$
        &\hspace{0.4cm}\begin{minipage}[c]{2.5cm}
                \epsfig{file=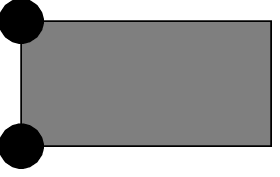,width=1.5cm,angle=270,
                bbllx=-5,bblly=0,bburx=55,bbury=85}
        \end{minipage}
        &\hspace{0.4cm}\begin{minipage}[c]{2.5cm}
                \vbox{\begin{center}
                \epsfig{file=table_1_2.eps,width=1.5cm,angle=270,
                bbllx=-20,bblly=0,bburx=55,bbury=85} 

                \hspace{-0.2cm}$p$
                \end{center}}
        \end{minipage}
        \hspace{0.4cm}\begin{minipage}[c]{2.5cm}
                \vbox{\begin{center}
                \epsfig{file=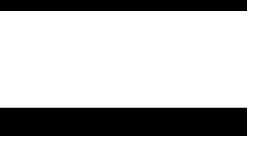,width=1.5cm,angle=270,
                bbllx=-20,bblly=0,bburx=55,bbury=85} 

                \hspace{-0.2cm}$1-p$
                \end{center}}
        \end{minipage}
        \\
        \hline
        $\sinh\left(\frac{\epsilon}{2}\right)$
        &\hspace{0.4cm}\begin{minipage}[c]{2.5cm}
                \epsfig{file=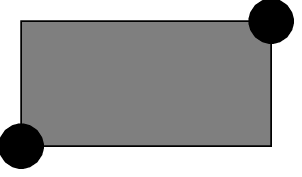,width=1.5cm,angle=270,
                bbllx=-5,bblly=0,bburx=55,bbury=85}
        \end{minipage}
        &\hspace{0.4cm}\begin{minipage}[c]{2.5cm}
                \vbox{\begin{center}
                \epsfig{file=table_3_2.eps,width=1.5cm,angle=270,
                bbllx=-20,bblly=0,bburx=55,bbury=85} 

                \hspace{-0.2cm}1
                \end{center}}
        \end{minipage}
        \\
        \hline   
    \end{tabular}
\caption{\it Cluster break-ups of various plaquette configurations together 
with their probabilities, where $p=2/[1+\exp(\epsilon/2)]$. The dots represent
occupied sites and the fat lines are the cluster connections.}
\end{center}
\end{table}

We use the same algorithm that was used in \cite{Cha99b}. Each configuration is 
decomposed into a set of clusters, which consist of connected lattice sites. A
new configuration is generated by flipping the clusters. When a cluster is 
flipped, all lattice sites contained in that cluster change occupation number 
from $n(x,t)$ to $1 - n(x,t)$, i.e. the occupied sites become empty and the 
empty ones occupied. To build the clusters, a probabilistic choice is made in 
each space-time interaction plaquette 
$[n(x,t),n(x+\hat i,t),n(x,t+1),n(x+\hat i,t+1)]$ as to which neighboring
lattice sites are connected to one another. A cluster is a sequence of
connected sites. In this algorithm, the clusters are closed loops. The 
probabilistic choices (called cluster break-ups) which build the clusters are 
designed to obey detailed balance and we only allow break-ups which generate
legal plaquettes under cluster flips. The cluster rules are illustrated in
Table 1. For plaquette configurations $[0,0,0,0]$ and $[1,1,1,1]$,
i.e. entirely empty or entirely occupied, we always connect sites with their 
time-like neighbors. For configurations $[1,0,0,1]$ and $[0,1,1,0]$ where a 
fermion hops to a neighboring site, we always connect sites with their
space-like neighbors. For configurations $[1,0,1,0]$ and $[0,1,0,1]$, i.e. a 
static fermion next to an empty site, we connect the sites with their
time-like neighbors with probability $p=2/[1+\exp(\epsilon/2)]$ and with their
space-like neighbors with probability $1-p$. This algorithm was also used in 
\cite{Wie94}. It is extremely efficient, has almost no detectable
autocorrelations and its dynamical exponent for critical slowing down is 
compatible with zero.
 
Each cluster has two orientations, with lattice site occupancies $n(x,t)$ and
$1 - n(x,t)$. When a cluster is flipped, the new configuration which is
generated may have a different sign from the previous one, depending on
whether or not the permutation of fermion world-lines is changed. A cluster
whose flip changes $\Sign[n]$ we call a meron, those which leave $\Sign[n]$
unchanged we call non-merons. Flipping a meron changes the topology of the
fermion world-lines. The term meron has been used before to denote
half-instantons \cite{Cal77}, such as in the 2-d $O(3)$ model at non-zero
vacuum angle $\theta$ \cite{Bie95}. The number of merons in a configuration is
always even, as flipping all clusters leaves the sign unchanged.
An example of a meron-cluster is given in Figure 1. When the meron-cluster is
flipped the first configuration with $\Sign[n] = 1$ turns into the second
configuration with $\Sign[n] = - 1$. For cluster algorithms more general than
the one described here, it is not always possible to identify certain clusters
as merons \cite{Cha00}.
\begin{figure}[t]
\hbox{
\hspace{2.7cm}
${\rm Sign[n]}=1$
\hspace{5.0cm}
${\rm Sign[n]}=-1$
}  
\begin{center}
\hbox{
\epsfig{file=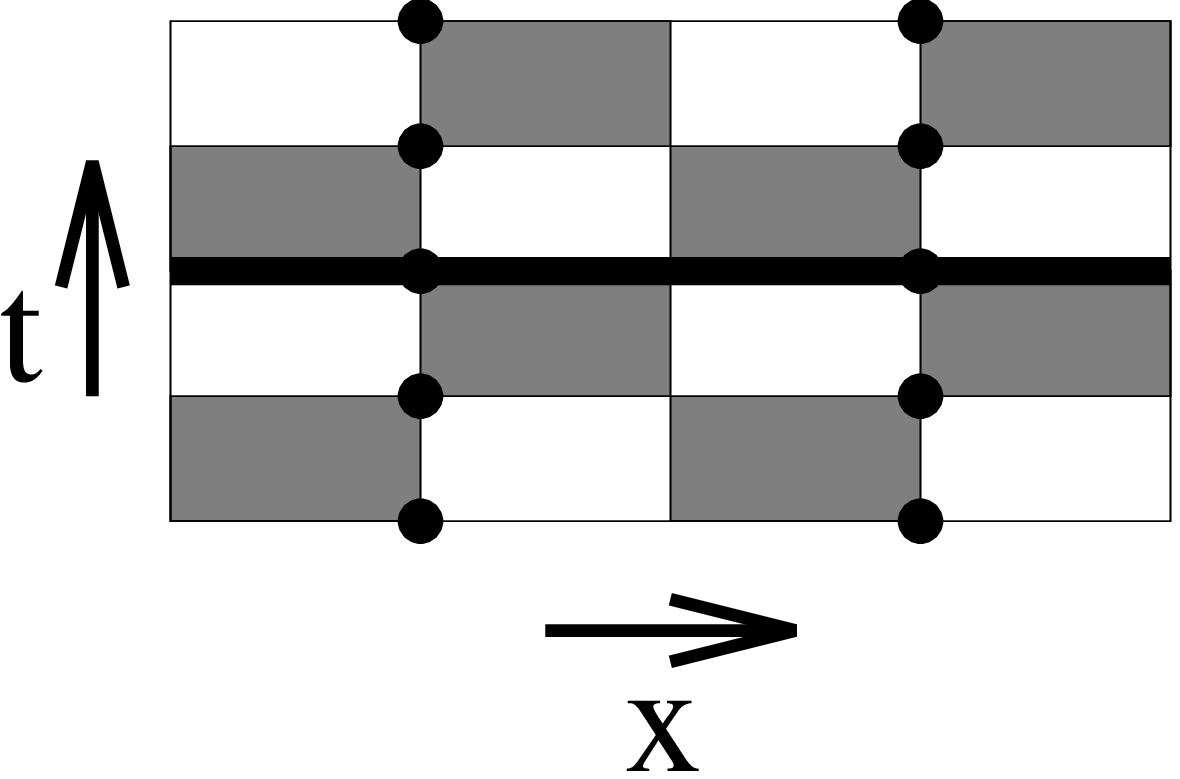,
width=4.5cm,angle=270,
bbllx=0,bblly=0,bburx=225,bbury=337}
\hspace{1.0cm}
\epsfig{file=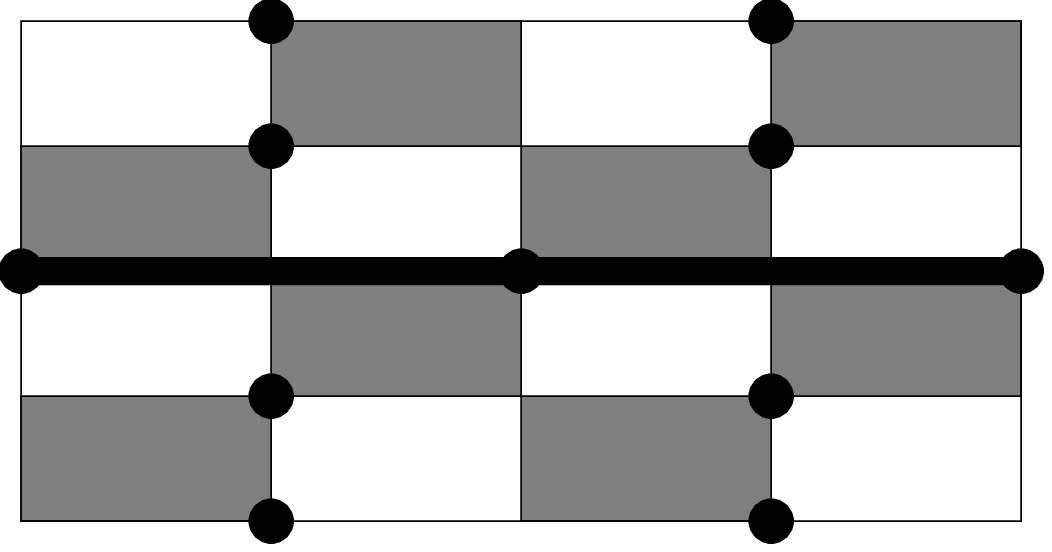,
width=4.5cm,angle=270,
bbllx=0,bblly=0,bburx=225,bbury=337}
}  
\end{center}
\caption{\it Two configurations of fermion occupation numbers in $(1+1)$
dimensions. The dots represent occupied sites and the shaded plaquettes carry
the interaction. With periodic spatial boundary conditions, the two fermions 
interchange positions in the second configuration, giving it $\Sign[n]=-1$. 
Flipping the meron-cluster (represented by the fat line) changes one 
configuration into the other, changing the fermion sign.}
\end{figure}

The meron concept alone gives us an exponential gain in statistics. Starting
from a configuration containing $N_C$ clusters, we consider the ensemble of 
$2^{N_C}$ configurations where we allow all possible cluster orientations. If
a configuration contains no merons, all configurations in the ensemble have 
$\Sign[n]=1$. However, if it contains merons, half the ensemble has
$\Sign[n]=1$ and the other half $\Sign[n]=-1$, which exactly cancel, giving a
contribution 0. The improved estimator gives $\langle \Sign \rangle=\langle 
\delta_{N,0} \rangle$, i.e. the probability that a configuration contains 
$N=0$ merons, which is an exponential improvement on standard algorithms,
which measure a statistical average of $\pm 1$. As explained in \cite{Cha99b},
this solves half of the sign problem.

We also construct improved estimators for observables. The chiral
susceptibility is
\begin{equation}
\chi = \frac{1}{\beta V} \langle (\PP)^2 \rangle_f = \frac{1}{\beta V}
\frac{\langle (\PP)^2 \Sign \rangle}{\langle \Sign \rangle}.
\end{equation}
The total chiral condensate for a given configuration, $\PP[n] = \sum_C
\PP_C$, is a sum of cluster contributions. Averaging $\chi$ over the ensemble 
of $2^{N_C}$ configurations gives
\begin{equation}
\label{chi}
\chi = \frac{\langle \sum_C |\PP_C|^2 \delta_{N,0} + 2 |\PP_{C_1}||\PP_{C_2}| 
\delta_{N,2} \rangle}{\beta V \langle \delta_{N,0} \rangle},
\end{equation}
This only gets contributions from configurations with $N=0$ or $N=2$ merons 
($C_1$ and $C_2$ are the two merons). The vast majority of configurations
contain many merons, but they make {\em no} contribution to observables. The
zero- and two-meron sectors of configuration space are exponentially small,
but they contain {\em all} of the contributions to $\chi$. Restricting
ourselves to only explore this sub-space, we exponentially enhance both the 
numerator and denominator of eq.(\ref{chi}), leaving the ratio
invariant. This solves the remaining half of the sign problem.

For the Binder cumulant $U$, we need to measure $\langle (\PP)^4 \rangle_f$ and hence
\begin{equation}
\langle \Sign (\PP)^4 \rangle = \langle \Sign \sum_{C_i,C_j,C_k,C_l} \PP_{C_i}
\PP_{C_j} \PP_{C_k} \PP_{C_l} \rangle.
\end{equation}
A cluster's condensate contribution $\PP_C$ changes sign when the cluster is flipped.
When a meron-cluster is flipped, $\Sign$ is changed. The non-zero terms in 
$\langle \Sign (\PP)^4 \rangle$ do not change sign if any cluster in the 
configuration is flipped. These non-zero terms must contain odd powers of 
$\PP_C$ for all merons $C$ in the configuration and even powers of $\PP_{C'}$
for all non-merons $C'$. The average over the ensemble of $2^{N_C}$
configurations is
\begin{eqnarray}
\label{u_est}
&& \langle \Sign (\PP)^4 \rangle_{2^{N_C}} = \delta_{N,0} \left[ \sum_C |\PP_C|^4
+ 6 \sum_{C,C'} |\PP_{C}|^2 |\PP_{C'}|^2 \right] \\ \nonumber
&& + \delta_{N,2} \left[ 4 |\PP_{C_1}|^3 |\PP_{C_2}| + 4 |\PP_{C_2}|^3 |\PP_{C_1
}|
+ 12 \sum_{C} |\PP_{C}|^2 |\PP_{C_1}| |\PP_{C_2}| \right] \\ \nonumber
&& + \delta_{N,4} \left[ 24 |\PP_{C_1}| |\PP_{C_2}| |\PP_{C_3}||\PP_{C_4}| \right],
\end{eqnarray}
where $N$ is the number of merons in the configurations, $C_1,C_2,C_3$ and
$C_4$ are the merons and all sums in eq.(\ref{u_est}) are over non-meron
clusters. This average only gets contributions from the zero-, two- and
four-meron sectors and so we need only explore this sub-space. We average this
quantity over the complete bosonic ensemble to measure $\langle \Sign (\PP)^4
\rangle$ and hence $U$.

Consider the case of measuring $\chi$. We expect that $p(0)/p(2) \propto 
(|C|/V \beta)^2$, where $p(0)$ and $p(2)$ are the probabilities that a 
configuration has zero or two merons and $|C|$ is the average cluster size. In
large volumes, the majority of configurations has two merons, contributing $0$
to $\langle \Sign \rangle$. For even greater accuracy, we reweight the 
meron-sectors with trial probabilities $p_t(0)$ and $p_t(2)$, so that they
appear with roughly equal frequency. This gives
\begin{equation}
\chi = \frac{\langle \sum_C |\PP_C|^2 \delta_{N,0} \ p_t(0) + 
2 |\PP_{C_1}||\PP_{C_2}| \delta_{N,2} \ p_t(2) \rangle}
{\beta V \langle \delta_{N,0} \ p_t(0) \rangle}.
\end{equation}
The reweighting probabilities can be adjusted to minimize the statistical
error. This technique was previously used in \cite{Bie95}. To measure the
Binder cumulant $U$, we use reweighting probabilities $p_t(0),p_t(2)$ and $p_t(4)$.
 
\section{Numerical Results}

\begin{table}[p]
\begin{center}
\begin{tabular}{|c|c|c|c|c|c|c|}
\hline
$L$ & $\beta$ & $\langle \Sign \rangle$ & $\chi$ & $p_t(0)/p_t(2)$ & 
$\langle \Sign \rangle_r$ & $\chi_r$ \\
\hline
\hline
  8 & 1.0 & 0.804(5) & 0.829(3) & 0.5/0.5 & 0.820(7) & 0.826(6) \\
\hline
  8 & 1.5 & 0.465(9) & 2.84(3) & 0.5/0.5 & 0.52(1) & 2.84(4) \\
\hline
  8 & 2.0 & 0.214(6) & 9.2(2) & 0.3/0.7 & 0.474(9) & 9.0(1) \\
\hline
  8 & 2.4 & 0.140(4) & 16.6(3) & 0.2/0.8 & 0.501(9) & 16.4(3) \\
\hline
 10 & 2.4 & 0.057(3) & 24.8(6) & 0.2/0.8 & 0.369(8) & 24.2(5) \\
\hline
 12 & 2.4 & 0.0203(8) & 33(1) & 0.1/0.9 & 0.443(7) & 34.0(7) \\
\hline
 14 & 2.4 & 0.0052(6) & 41(4) & 0.1/0.9 & 0.338(8) & 44(1) \\
\hline
 16 & 2.4 & 0.0005(2) & 80(40) & 0.075/0.925 & 0.314(4) & 57(1) \\
\hline
 20 & 2.4 & --- & --- & 0.03/0.97 & 0.355(9) & 82(3) \\
\hline
 24 & 2.4 & --- & --- & 0.01/0.99 & 0.46(1) & 120(5) \\
\hline
 28 & 2.4 & --- & --- & 0.01/0.99 & 0.329(9) & 156(8) \\
\hline
\end{tabular}
\end{center}
\caption{\it Numerical results for the non-reweighted $\langle \Sign \rangle$
and susceptibility $\chi$ measured over all meron sectors, and the reweighted 
$\langle \Sign \rangle_r$ and $\chi_r$ measured over the zero- and two-meron
sectors only with a reweighting factor $p_t(0)/p_t(2)$. For the larger
volumes, $\langle \Sign \rangle$ and $\chi$ cannot be measured. }
\end{table}

\begin{figure}[p]
\label{histograms}
\vbox{
\begin{center}
\psfig{figure=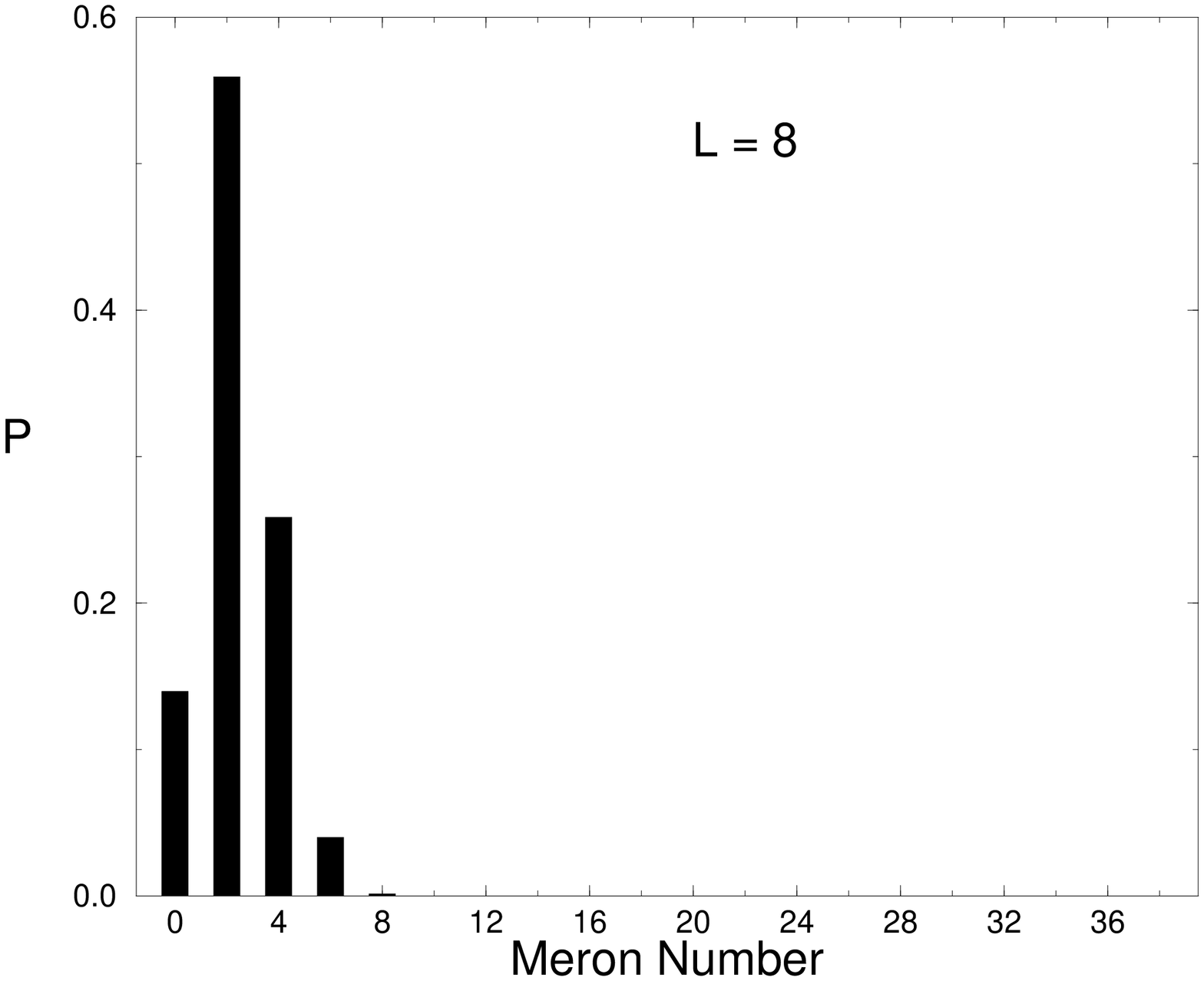,height=2.5in,width=2in,angle=270}
\psfig{figure=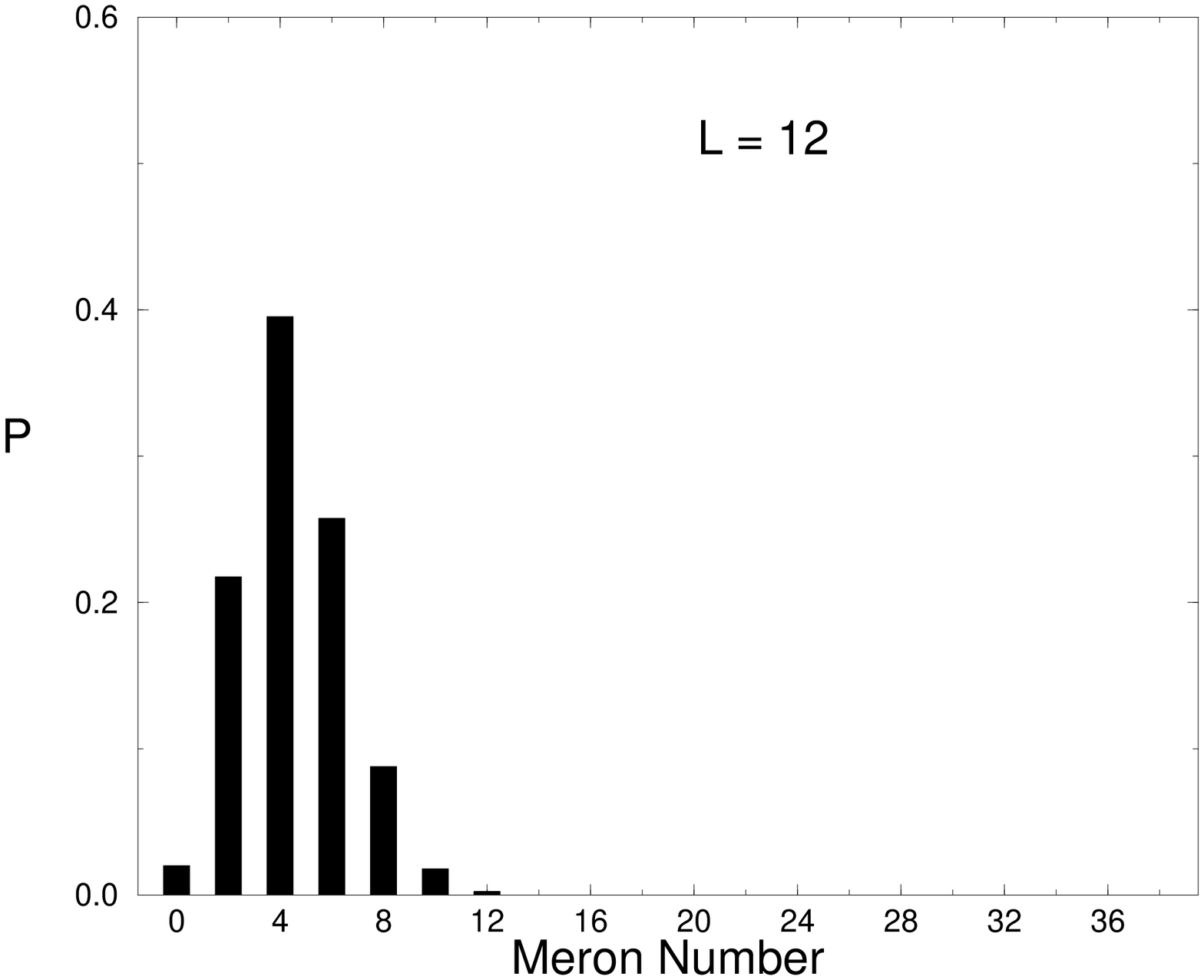,height=2.5in,width=2in,angle=270}
\psfig{figure=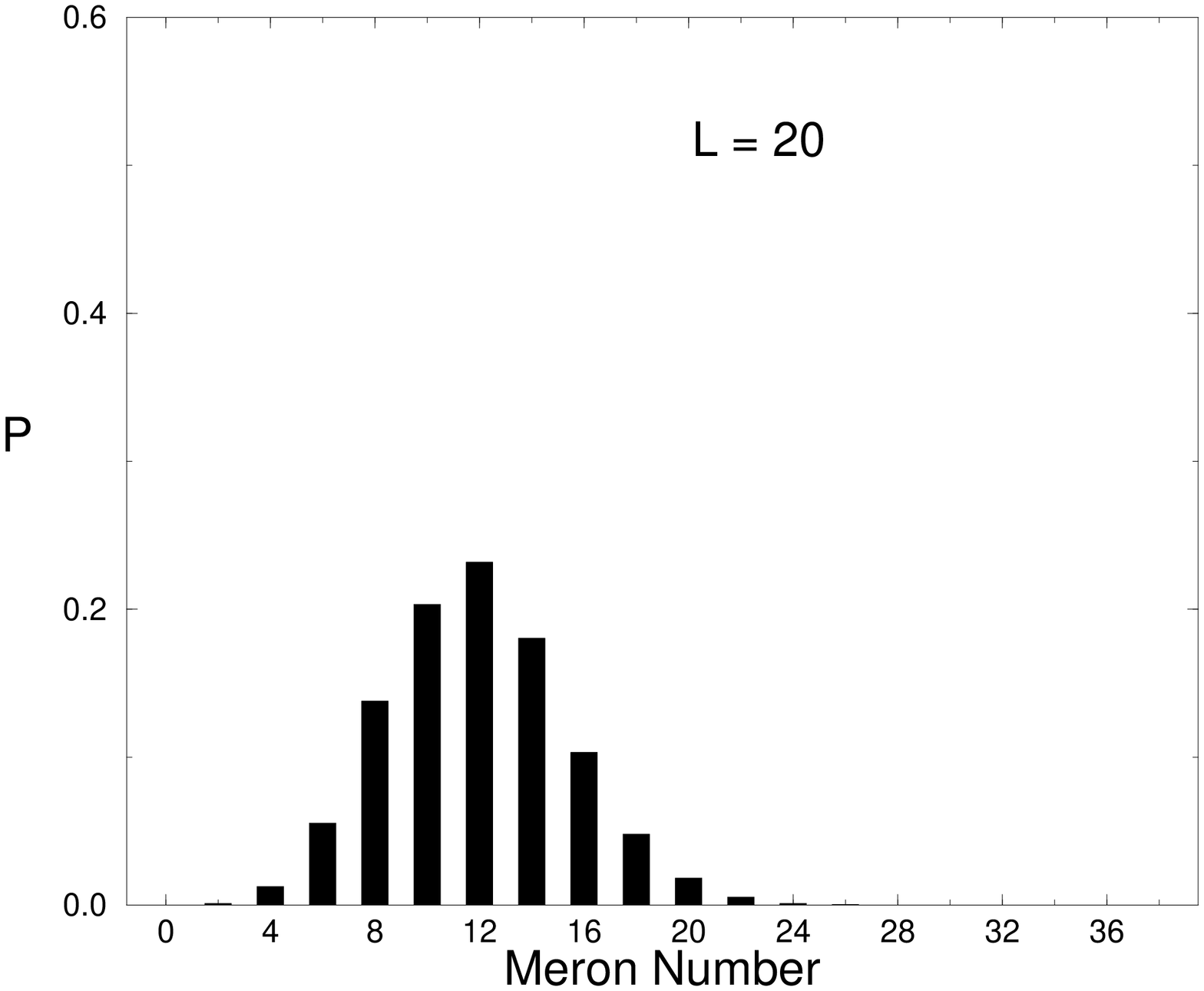,height=2.5in,width=2in,angle=270}
\psfig{figure=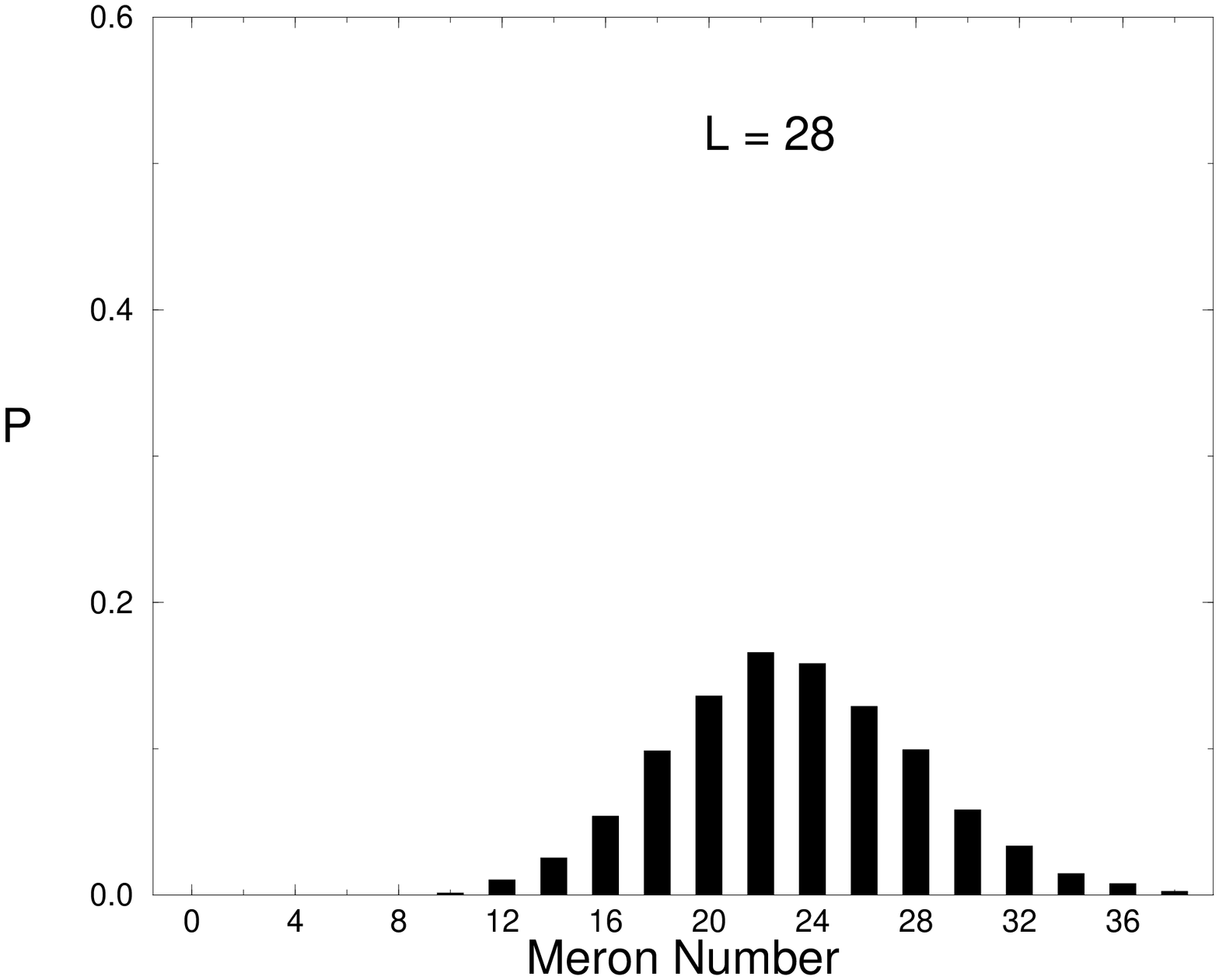,height=2.5in,width=2in,angle=270}
\caption{\it The meron number probability distribution for various
spatial sizes $L = 8,12,20$ and $28$ at $\beta = 2.4$.}
\end{center}
}
\end{figure}

We have performed simulations of the staggered fermion model on lattices with
antiperiodic spatial boundary conditions from $L = 4$ up to $L = 30$ at
inverse temperatures in the range $\beta \in [1.0,3.0]$, which includes the 
critical temperature where the chiral symmetry is spontaneously broken. 
We have made separate runs with either a fixed number of time slices 
(typically $M=10$, i.e. 40 time slices) or with fixed lattice spacing in the
time direction ($\epsilon = 0.1$). In each simulation, we have
made at least 1000 thermalization sweeps followed by 10000 measurements, with
these numbers increased by a factor of 10 for $L \leq 10$. In one sweep of the
lattice, a new cluster connection is proposed on each interaction plaquette
and each cluster is flipped with probability $1/2$. To find the optimal
reweighting probabilities $p_t(N)$ which minimize the statistical error, we
first make a sample run without reweighting, only exploring the relevant 
meron-sectors. The observed relative weights are then used in production runs,
where the sectors appear with equal probability. The major part of the sign
problem is removed by the improved estimators, but reweighting is necessary
for accurate measurements in large volumes.

\begin{figure}[thb]
\begin{center}
\psfig{figure=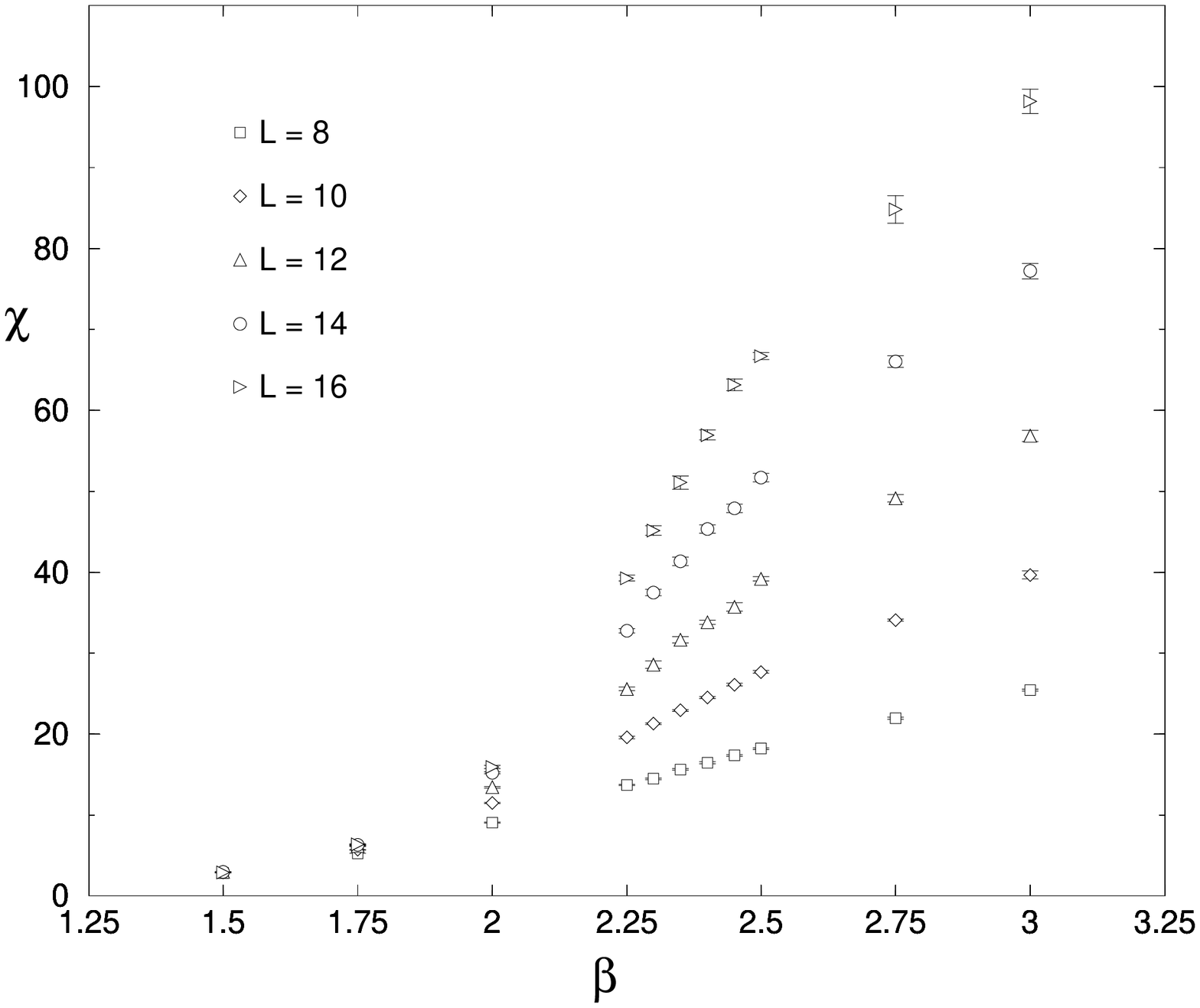,height=4in,width=3.2in,angle=270}
\caption{\it The chiral susceptibility $\chi$ as a function of the inverse 
temperature $\beta$ for various spatial sizes $L = 8,10,12,14$ and $16$ on
lattices with $40$ time slices. The chiral symmetry is intact at small $\beta$
and spontaneously broken at large $\beta$.}
\end{center}
\end{figure}

A sample of the data measured is given in Table 2. The Table contains $\langle
\Sign \rangle$ and the susceptibility $\chi$ measured over all meron-sectors, 
and the reweighted $\langle \Sign \rangle_r$ and $\chi_r$ measured over the 
zero- and two-meron sectors only with the reweighting factor $p_t(0)/p_t(2)$.
All of these data are produced with 1000 thermalization sweeps and 10000
measurements. As all of the contributions to $\chi$ come from the zero- and 
two-meron sectors, $\chi$ and $\chi_r$ should be identical. Note that 
$\langle \Sign \rangle_r$, the fraction of zero-meron configurations generated
by sampling the zero- and two-meron sectors only, is typically a lot bigger
than $\langle \Sign \rangle$, the fraction of zero-meron configurations generated
over all meron sectors. In small space-time volumes, $\chi$ can be accurately 
measured even when sampling all meron sectors. However, in large space-time 
volumes, $\langle \Sign \rangle$ is too small to be measured and we can only 
determine the susceptibility by restricting ourselves to the zero- and
two-meron sectors. The staggered fermion model suffers from a very severe sign
problem which is solved by the Meron-Cluster algorithm.

Figure 2 shows the meron number probability distribution in an 
algorithm that samples all meron sectors without reweighting. For small volumes
the zero-meron sector and hence $\langle \Sign \rangle$ are relatively large, 
while multi-meron configurations are rare. On the other hand, in larger volumes
the vast majority of configurations has a large number of merons and hence 
$\langle \Sign \rangle$ is exponentially small. For example, an extrapolation 
from smaller volumes gives a rough estimate for the non-reweighted 
$\langle \Sign \rangle \approx 10^{-9}$ on the $L=28$ lattice at 
$\beta = 2.4$., while the reweighted $\langle \Sign \rangle_r = 0.329(9)$. 
Even if the configurations are entirely uncorrelated, to achieve a similar 
accuracy without the meron-cluster algorithm one would have to increase the 
statistics by a factor $10^{18}$, which is obviously impossible. In fact, at 
present there is no other method that can be used to simulate this model.

\begin{figure}[thb]
\label{chi_vs_l}
\begin{center}
\psfig{figure=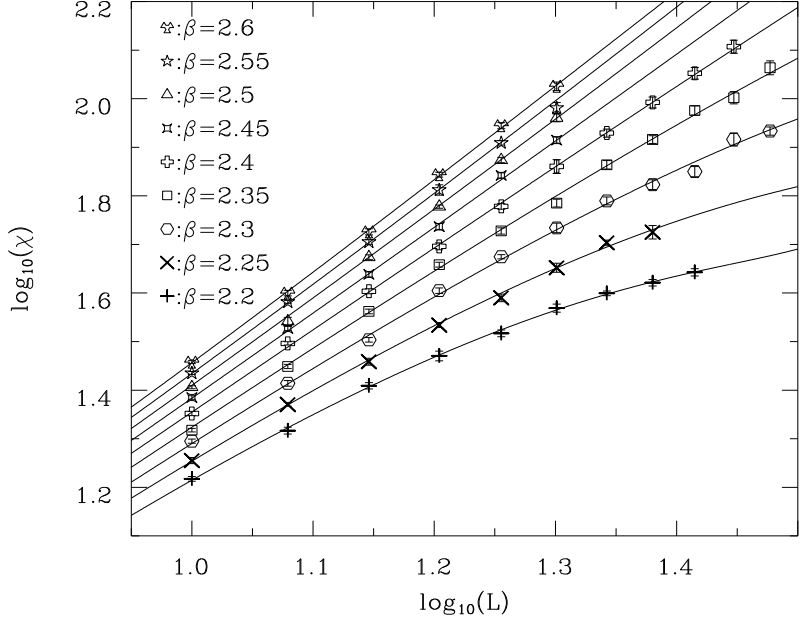,height=6in,width=4in,angle=90}
\caption{\it The chiral susceptibility plotted against $L$ for various
values of $\beta$ computed with $\epsilon=0.1$. The fit is to the
finite-size scaling formula (\ref{scaling}) with $a(x)$ expanded to
first order and $b(y)$ expanded to third order. The exponents are
set to the 2-d Ising model values. All curves are obtained from
one fit. The $\chi^2$ per degree of freedom is 0.84 indicating a good agreement of
our data with the finite-size scaling ansatz and 2-d Ising model 
critical exponents.}
\end{center}
\end{figure}

Figure 3 shows the chiral susceptibility $\chi$ as a function of 
$\beta$ for various spatial sizes $L$. At high temperatures (small $\beta$) 
$\chi$ is almost independent of the volume, indicating that chiral symmetry is
intact. On the other hand, at low temperatures (large $\beta$) $\chi$
increases with the volume, which implies that chiral symmetry is spontaneously
broken. To study the critical behavior in detail, we have performed a finite-size
scaling analysis for $\chi$ focusing on the range $\beta \in [2.2,2.6]$
around the critical point. Since a $\Z(2)$ chiral symmetry is spontaneously 
broken at finite temperature in this $(2+1)$-d model, one expects to find the 
critical behavior of the 2-d Ising model. The corresponding finite-size scaling
formula valid close to $\beta_c$ is \cite{Blo95}
\begin{eqnarray}
\label{scaling}
&&\chi(L,\beta) = a(x) + b(y) L^{\gamma/\nu}, 
\nonumber \\
&&a(x) = a_0 + a_1 x + a_2 x^2 + ..., \ x = \beta - \beta_c,
\nonumber \\
&&b(y) = b_0 + b_1 y + b_2 y^2 + ..., \ y = (\beta - \beta_c) L^{1/\nu}.
\end{eqnarray}
For the 2-d Ising model the critical exponents are given by $\nu = 1.0$ 
and $\gamma/\nu = 1.75$. Assuming these values for the exponents, we obtain
$\beta_c=2.43(1)$ for fixed $\epsilon=0.1$ from the finite-size scaling fit,
with a chi squared per degree of freedom of 0.84. The fit of the data is 
plotted in Figure 4. The value of $\beta_c$ is slightly dependent on $\epsilon$.

\begin{figure}[thb]
\label{uni_curve}
\begin{center}
\psfig{figure=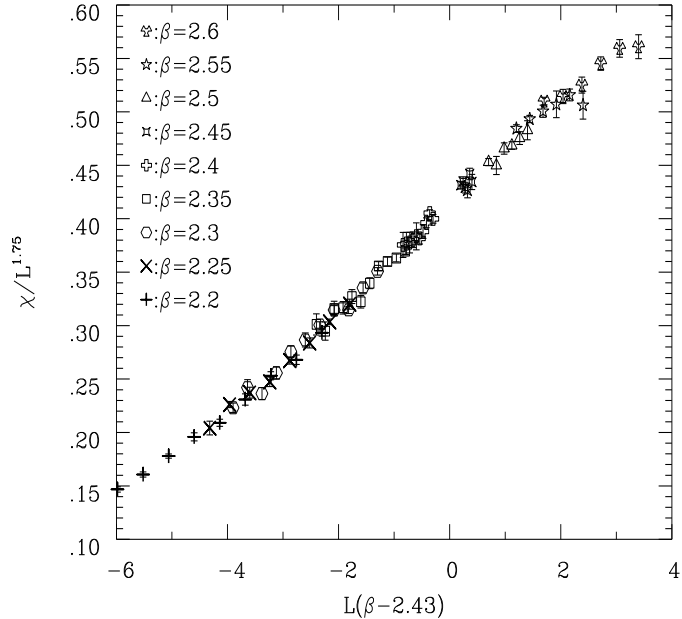,height=5in,width=4in,angle=90}
\caption{\it Finite-size scaling behavior of the chiral susceptibility $\chi$. 
The data for various spatial sizes $L$ and inverse temperatures $\beta$ fall on one 
universal curve.}
\end{center}
\end{figure}

%\begin{figure}[p]
%\begin{center}
%\psfig{figure=chi_bc.eps,height=3in,width=4.5in,angle=270}
%\caption{\it Finite-size scaling behavior of the chiral susceptibility $\chi$ 
%as a function of the spatial size $L$ at the estimated critical inverse 
%temperature $\beta_c$. A fit of the volume-dependence (dashed line) 
%gives the critical exponent $\gamma/\nu = 1.745(5)$ of the 2-d Ising model.}
%\end{center}
%\end{figure}

In the finite-size scaling equation (\ref{scaling}), for large enough $L$ one can 
neglect the term $a(x)$. Then $\chi/L^{\gamma/\nu}$ is a function of $y = 
(\beta - \beta_c) L^{1/\nu}$ alone, i.e. the susceptibility data in various
volumes at various $\beta$ can be described by one universal function. We have
varied the value $\beta_c$ to find if all the data can be collapsed onto one 
universal curve. In Figure 5, we plot the universal curve obtained by taking 
$\beta_c=2.43$. The excellent agreement over a large range of spatial volumes
$L$ and inverse temperatures $\beta$ is an indication of the quality of the
finite-size scaling fit. 

\begin{figure}[p]
\label{binder_theory}
\begin{center}
\psfig{figure=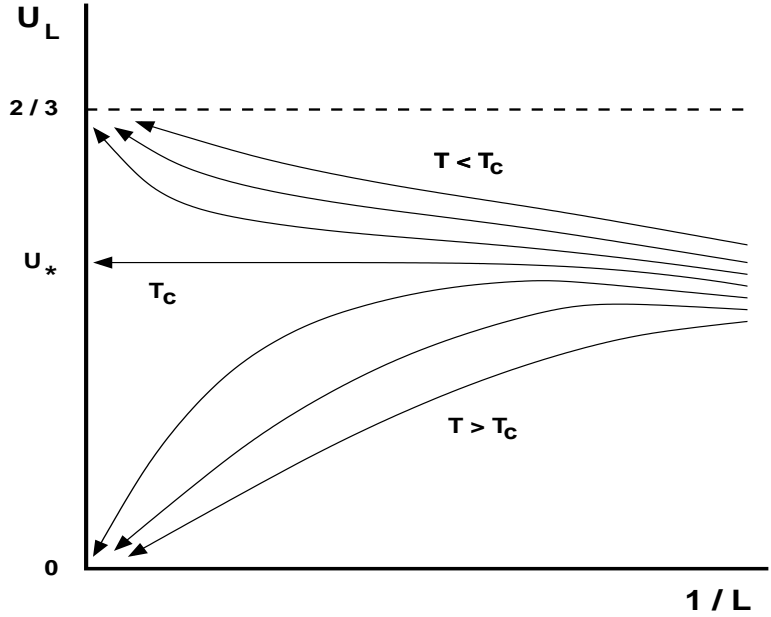,height=3.2in,width=4in,angle=0}
\caption{\it The expected flow of $U_L$ as a function of $1/L$. On each curve
the temperature is constant.}
\end{center}
\end{figure}

\begin{figure}[p]
\label{binder_actual}
\begin{center}
\psfig{figure=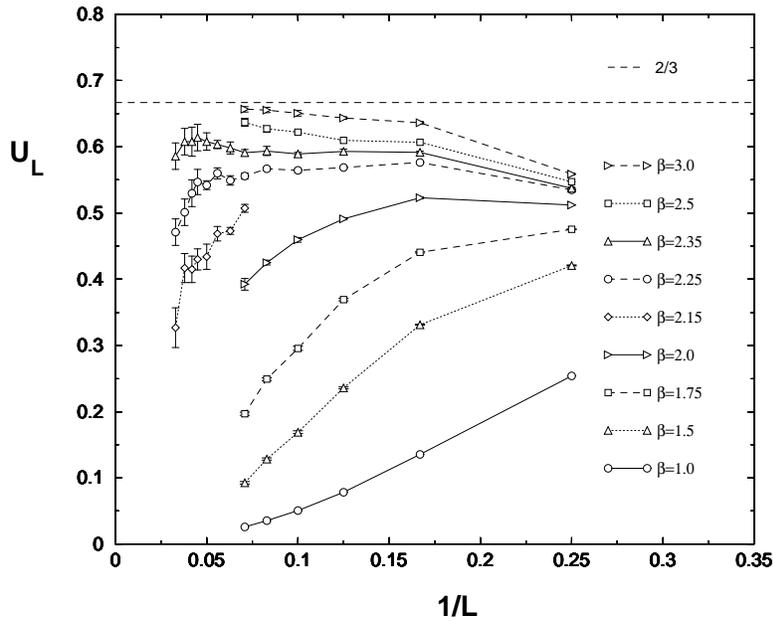,height=4in,width=3.2in,angle=270}
\caption{\it Measured values of $U_L$ plotted versus $1/L$ for various $\beta$
in volumes with $40$ time slices. Near $\beta=2.35$, the values appear to flow
into the non-trivial fixed point $U_*=0.60(1)$.}
\end{center}
\end{figure}

We also measure $U_L$, the Binder cumulant in volumes of extent $L$. In Figure
6, we plot the expected behavior of $U_L$ as $L$ increases for different
temperatures. For $T>T_c$, the chiral symmetry is intact and $U_L$ flows into
the $T=\infty$ fixed point $U=0$. For $T<T_c$, the chiral symmetry is
spontaneously broken and $U_L$ flows into the $T=0$ fixed point $U=2/3$. If
the universality class has a non-trivial fixed point $U=U_*$, then $U_L$ flows
into this value at $T=T_c$. By measuring $U_L$ in various volumes at many
different temperatures, we determine this flow numerically. We have measured
the Binder cumulant values in volumes up to $L=30$ and we plot some of these
values as a function of $1/L$ in Figure 7. These measurements are made with
the number of time slices fixed at $40$. Each curve in the figure represents
some fixed temperature. In Figure 7, for small $\beta$ (i.e. high
temperatures), $U_L$ clearly flows into the infinite temperature fixed point 
$U=0$, while for $\beta$ large (low temperatures), $U_L$ flows into the zero 
temperature fixed point $U=2/3$. For $\beta$ close to $\beta_c$, we have to go
to larger volumes to see this behavior. Near $\beta=2.35$, the cumulant values 
appear to flow into a non-trivial fixed point $U_*$. Examining this region 
closely, we estimate the critical inverse temperature as $\beta_c=2.36(2)$ and
the fixed point value $U_*=0.60(1)$. The finite-size scaling fit of $\chi$
measured at this $\epsilon \approx 0.24$ gives the same value of $\beta_c$.
Note that this deviates slightly from the critical temperature measured at 
$\epsilon = 0.1$. The universal fixed point value for the 
2-d Ising model is estimated as $U_* \sim 0.58$ \cite{Bin81}. This is 
further evidence that the chiral phase transition belongs to the 2-d Ising 
universality class.

\section{Conclusions}

The Meron-Cluster algorithm has recently been developed to allow numerical
simulations in models which suffer from a very severe sign problem. In this
paper, we have applied this technique to investigate a model of staggered 
fermions. Unlike standard methods, which integrate out the fermions, 
resulting in a non-local bosonic action, we use a Fock space of occupation
number to describe the fermions. We have a local bosonic action, with an 
additional non-local sign factor which contains the Fermi statistics. Due
to the Pauli exclusion principle, configurations which have an odd 
permutation of fermion world lines have a negative sign. This sign leads to
very large cancellations in observables and usually makes it impossible to make
accurate measurements in numerical simulations. The Meron-Cluster algorithm
decomposes every configuration into closed loops of connected sites, each
loop being a cluster. Loops which change the fermion sign when flipped are
identified as meron-clusters. A meron-cluster identifies a pair of
configurations with equal weight and opposite sign. This results in an exact 
cancellation of two contributions $\pm 1$ to the path integral, such that only
configurations without merons contribute to the partition
function. Observables only receive contributions from configurations which
contain very few or no merons, whereas the vast majority of configurations
contain many merons. By only exploring the sectors of configuration space
with the relevant numbers of merons, one makes an exponential gain in
statistics. Combined with efficient re-weighting of the remaining meron
sectors, this completely solves the sign problem. Cluster algorithms are
extremely efficient at exploring configuration spaces and generating uncorrelated
configurations and generally do not suffer from critical slowing down. Even in
models without a sign problem, the Meron-Cluster algorithm is more efficient 
than standard fermion simulation methods.

In this paper, we examined a model of $N=1$ flavor of staggered fermions in 
$(2+1)$-dimensions, which has a $\Z(2)$ chiral symmetry. The model has a very
severe sign problem and cannot be solved by standard fermion simulation
algorithms. Using a Meron-Cluster algorithm, we were able to make 
high-precision measurements of the chiral susceptibility and Binder cumulant
even in very large volumes and low temperatures. In order to perform an
accurate and reliable finite-size scaling analysis, it was necessary to go to
volumes so large, where the sign problem is so severe, that a standard
algorithm would require statistics on the order of $10^{18}$ to attain a
similar accuracy. We were able to verify that the model undergoes a 
finite-temperature chiral phase transition, which belongs to the universality
class of the $2-$d Ising model. This is the behavior expected from dimensional
reduction and universality. The same universal behavior was observed in the 
$N = 4$ flavor case \cite{Kog99}. However, the standard fermion algorithm that
was used in that study does not work for $N < 4$ due to the fermion sign problem.

It is quite natural to use cluster algorithms in models of discrete variables.
A future possible application of the Meron-Cluster algorithm is in exploring
quantum link models \cite{Cha97} which are used in the D-theory formulation of QCD 
\cite{Bro97,Bea98,Wie99}. In D-theory, a model of discrete quantum variables
undergoes dimensional reduction, resulting in an effective theory of
continuous classical variables. In quantum link QCD the quarks arise as domain
wall fermions. The application of meron-cluster algorithms to domain wall
fermions is in progress. Also there are many applications to sign problems in 
condensed matter physics. Investigations of antiferromagnets in a magnetic
field and of systems in the Hubbard model family are given in Ref.\cite{Cox99}.

At present, the Meron-Cluster algorithm is the only method that allows us to
solve the fermion sign problem. A severe sign problem arises in lattice QCD
calculations at non-zero baryon number due to a complex action. It is therefore
natural to ask if our algorithm can be applied to this case. At non-zero 
chemical potential the 2-d $O(3)$ model, which is a toy model for QCD, also 
suffers from a sign problem due to a complex action. When applied to the 
D-theory formulation of this model, the Meron-Cluster algorithm solves the sign
problem completely \cite{Cox99}. It is an open question if such progress can
be made in investigations of QCD.

%-----------------------------------------------------------------------
% The lines below are necessary in order to enumerate the equations
% according to the sections where they are.
\makeatletter
\@addtoreset{equation}{section}
\makeatother
\renewcommand{\thesection}{\Alph{section}}
\setcounter{section}{0}
%-----------------------------------------------------------------------

\section*{Acknowledgements}

We would like to thank Shailesh Chandrasekharan and Uwe-Jens Wiese for helpful
discussions.

\end{document}